\documentclass[aps,prl,floatfix,twocolumn,showpacs,preprintnumbers,amsmath,amssymb,superscriptaddress]{revtex4-1}
\usepackage{amsmath}
\usepackage{amssymb}
\usepackage{graphicx}
\usepackage{hyperref} 
\usepackage{bm}
\usepackage{epstopdf}
\usepackage{epsfig}
\usepackage{comment}
\usepackage{color}

\begin{document}
\title{Threshold behavior in spin lasers: Spontaneous emission and nonlinear gain}
\author{Gaofeng Xu}
\email{xug@hdu.edu.cn}
\affiliation{Department of Physics, Hangzhou Dianzi University, Hangzhou, Zhejiang 310018, China}
\affiliation{Department of Physics, University at Buffalo, State University of New York, Buffalo, NY 14260, USA}
\author{Krish Patel}
\affiliation{Department of Physics, University at Buffalo, State University of New York, Buffalo, NY 14260, USA}
\author{Igor \v{Z}uti\'c}
\email{zigor@buffalo.edu}
\affiliation{Department of Physics, University at Buffalo, State University of New York, Buffalo, NY 14260, USA}
\begin{abstract}
A hallmark of spin-lasers, injected with spin-polarized carriers, is their threshold behavior with 
the onset of stimulated emission. 
Unlike the single threshold in conventional lasers with unpolarized carriers, two thresholds are expected in spin lasers. With the progress in scaled-down lasers and the use of novel two-dimensional materials it is unclear if the common description of spin lasers assuming a negligible spontaneous emission and linear optical gain remains relevant or even how to identify the lasing thresholds. Our rate-equation description addresses these questions by considering a large spontaneous emission and a nonlinear optical gain. We provide a transparent approach and analytical results to explore the resulting threshold behavior, its deviation from the prior studies, as well as  guide future spin-lasers as a versatile platform for spintronics beyond magnetoresistance. 
\end{abstract} 
\maketitle

The use of lasers in practical applications usually reflects their highly-controllable nonlinear coherent response~\cite{Chuang:2009, Coldren:2012,Michalzik:2013}. 
With a current injection required for the onset of lasing, there is a striking change from incoherent to coherent emitted light that can be described by the Landau theory of 
second-order phase transitions~\cite{Haken:1985,DeGiorgio1970:PRA}. However, in scaled-down
lasers, a large contribution of the spontaneous emission blurs this transition and even a thresholdless lasing is 
possible~\cite{Oulton2009:N,Khajavikha2012:N,Saxena2013:NP,Jagsch2018:NC,Bhattacharya2017:PRL}, while offering 
advantages for optical interconnects, imaging, sensing, and biological applications~\cite{Ma2019:NN}.

\begin{figure}[h]
\centering
\includegraphics*[width=8.6cm]{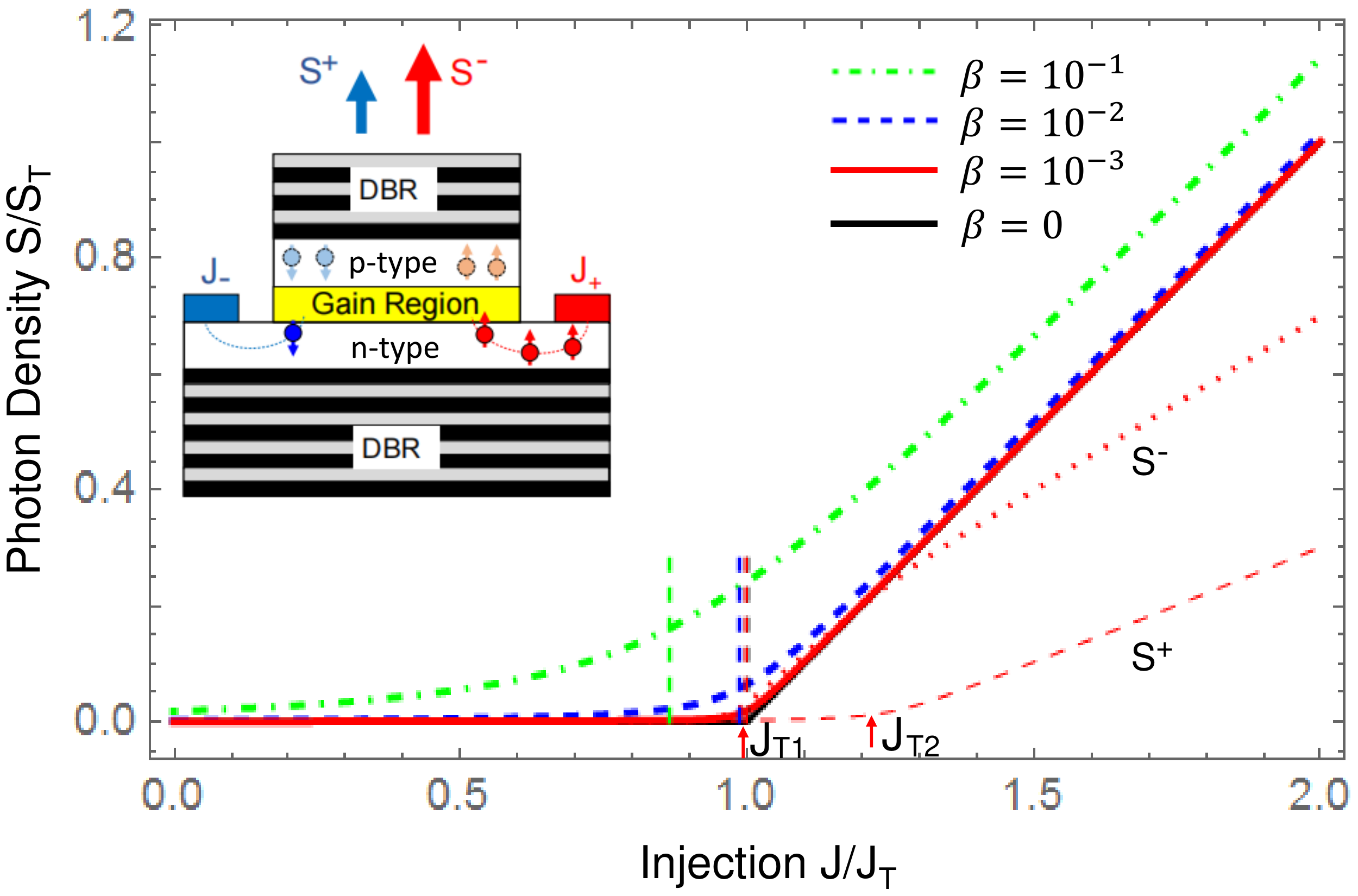} 
\vspace{-0.5cm}
\caption{Photon density 
$S$ as functions of injection $J$ for $\beta=0$, $10^{-3}$, $10^{-2}$, $10^{-1}$, with the vertical lines indicating the threshold injection $J_T(\beta)$, respectively. 
Helicity-resolved 
$S^{\pm}$ are shown as  a function 
of injection with polarization $P=0.2$ for $\beta=10^{-3}$ and the corresponding thresholds $J_{T1, 2}$ (arrows). The photon densities and injection are normalized by $J_T\equiv J_T(0)$ and $S_T=S(2 J_T)$. Inset: A schematic of a spin laser formed by a gain region, $p$- and $n$-type semiconductor layers, and distributed Bragg reflectors (DBR), with injection of different spins (here $J_+>J_-$, $J \equiv J_++J_-$) and circularly-polarized emission with $S^->S^+$.}
\label{fig:1}
\vspace{-0.3cm}
\end{figure}

Through conservation of angular momentum, injecting spin-polarized carriers in lasers 
offers control of the helicity of emitted light and room-temperature spintronic applications, 
beyond magnetoresistance~\cite{Zutic2004:RMP,Nishizawa2017:PNAS,Zutic2020:SSC,Rozhansky2021:PE,Mantsevich2019:PRB,%
Nishizawa2021:MM,Tsymbal:2019}. Such spin lasers~\cite{Hallstein1997:PRB,Ando1998:APL,Rudolph2003:APL,Rudolph2005:APL,Holub2007:PRL,%
Gerhardt2006:EL, Hovel2008:APL,Basu2009:PRL,Saha2010:PRB,Li2010:APL,Gerhardt2011:APL,Iba2011:APL,Iba2012:SSC,Frougier2013:APL,Cheng2014:NN,Alharthi2014:APL,Alharthi2015:APL,%
Lindemann2016:APL,Yokota2017:IEEEPTL,Yokota2018:APL,Lindemann2019:N}
can exceed the performance of the best conventional counterparts 
(with spin-unpolarized carriers) as well as motivate emerging device concepts, from spin interconnects~\cite{Dery2011:APL, Zutic2019:MT} to phonon lasers~\cite{Khaetskii2013:PRL}.
Typically, spin lasers are realized as  a vertical cavity surface emitting laser (VCSEL)~\cite{Michalzik:2013},
shown in the inset of Fig.~\ref{fig:1}. 

The rate equations for spin lasers~\cite{Rudolph2003:APL,Rudolph2005:APL,Holub2007:PRL,Gothgen2008:APL,Lee2010:APL,Lee2012:PRB, Lee2014:APL}
often rely on a linear gain model, common to bulk materials, despite the nonlinear gain dependence on the carrier density~\cite{Chow:1999, Holub2011:PRB, FariaJunior2017:PRB, FariaJunior2015:PRB}.
Since the carrier density dependence of the gain in systems of reduced dimensionality, such as quantum wells,  are more accurately described  by a logarithmic function~\cite{Chuang:2009, Coldren:2012,Michalzik:2013,FariaJunior2017:PRB},
here we consider usually overlooked effects of spontaneous emission and logarithmic gain on the threshold behavior of spin lasers. 
With a general gain model, the rate equations can be expressed as~\cite{Gothgen2008:APL, Lee2010:APL},
\begin{eqnarray}
\frac{dn_{\pm}}{dt} &=& J_{\pm}-g_{\pm}S^{\mp}-(n_{\pm}-n_{\mp})/\tau_s-R_{sp}^{\pm}, 
\label{eq:ren}
\\
\frac{dS^{\pm}}{dt} &=& \Gamma g_{\mp}S^{\pm}-S^{\pm}/\tau_{ph}+\beta \Gamma R_{sp}^{\mp}, 
\label{eq:reS}
\end{eqnarray}

\noindent{where} $n_{\pm}$ are the densities of spin-up (down) $+$ ($-$) electrons, with a spin-relaxation time $\tau_s$. 
Since holes have much shorter spin-relaxation times than those of electrons in 
semiconductor quantum wells~\cite{Zutic2004:RMP}, only the electrons are spin polarized, while charge neutrality yields the densities of holes $p_{\pm}=(n_+ + n_-)/2$. $J_{\pm}$ are the injection rates of spin-up (down) $+$ ($-$) electrons, with the polarization $P=(J_+ - J_-)/J$. $S^{\pm}$ are the photon densities of positive  ($+$) and negative ($-$) helicity.  
The spontaneous recombination of carriers can be expressed in a linear form as $R_{sp}^{\pm}=n_{\pm}/\tau_r$, characterized by a carrier recombination 
time $\tau_r$~\cite{Bourdon2002:JAP,Hader2005:SPIE,Gothgen2008:APL}. 
The optical gain can have various dependence on the carrier density.  In a simple linear  
model, the spin-dependent gain takes the form $g_{\pm}= g_0 (n_{\pm} + p_{\pm}-n_{tran})$, with $g_0$ the gain parameter and $n_{tran}$ the transparency carrier density. $\Gamma$ is the optical confinement factor and $\tau_{ph}$ is the photon lifetime in the cavity. The spontaneous emission factor $\beta$ characterizes the fraction of spontaneous emission 
coupled to the lasing mode. For a typical spin laser, $\beta \sim 10^{-5}-10^{-3}$~\cite{Rudolph2005:APL,Holub2007:PRL}, justifies the $\beta=0$ approximation.

The threshold behavior of (spin) lasers without significant contribution from spontaneous emission has been well established. 
When $\beta=0$, the steady-state solution of the rate equations without spin injection ($J_+=J_-$) gives the threshold carrier density $n_T=n_{tran} +1/(\Gamma  g_0 \tau_{ph})$ and steady-state photon density $S=\Gamma \tau_{ph} (J-n_T/\tau_r)$, which leads to the threshold injection $J_T= n_T/\tau_r$~\cite{Gothgen2008:APL}. 

However, the spontaneous emission is in general not negligible. 
In scaled-down lasers,  due to their reduced cavity sizes and the Purcell effect~\cite{Purcell1946:PR}, there are reports 
of significantly enhanced spontaneous emission factors up to $\beta\sim0.1$ , which  
significantly changes the threshold behavior and even thresholdless lasers are possible~\cite{Oulton2009:N,Khajavikha2012:N,%
Saxena2013:NP,Jagsch2018:NC,Bhattacharya2017:PRL}. 
To explore the effect of $\beta$ on the lasing threshold, we obtain an analytical relation between $S$ and $J$ by solving the steady-state rate equations with $\beta \neq 0$, 
\begin{widetext}
\begin{eqnarray}
S(J)  =\frac{1}{2}\Gamma \tau_{ph} (J  - n_T/\tau_r)  +\frac{  \sqrt{\left[\Gamma g_0 \tau_{ph} (J \tau_r -n_T)+ \Gamma g_0 \tau_{ph} n_T \beta -\beta\right]^2 + 4\Gamma g_0 \tau_{ph} \tau_r J \beta} +(\Gamma g_0 n_T \tau_{ph})\beta } {2g_0 \tau_r}. 
\label{Eq:SJ}
\end{eqnarray}
\end{widetext}
In Fig.~\ref{fig:1}  
we see that, with increasing $\beta$, the originally abrupt transition from nonlasing to lasing region gradually becomes smoother,  
which makes the definition of lasing threshold challenging. 

\begin{figure}[ht]
\centering
\includegraphics*[width=8.6cm]{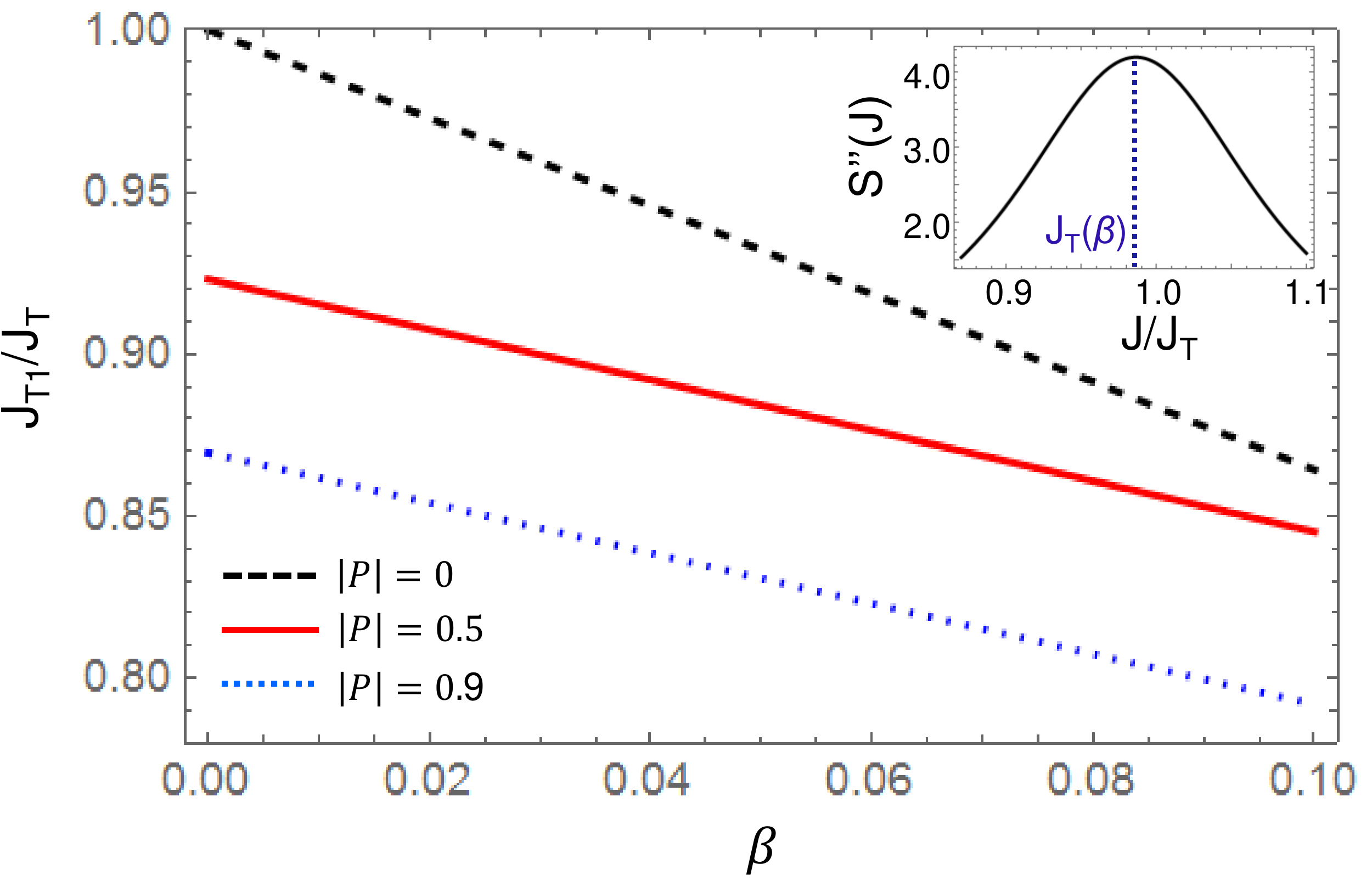}
\vspace{-0.5cm}
\caption{Dependence of $J_{T1}/J_T$ on $\beta$ for injection with polarization $|P|=0$, $0.5$, and $0.9$. $J_{T1}$ is normalized by its threshold $J_T$ at $P=0$ and $\beta=0$. Inset: the second derivative $S''(J)$ for $\beta=0.01$, with the peak position, i.e., threshold $J_T (\beta)$, marked by the vertical line. The photon density $S$ is normalized by $S_T$.}
\label{fig:JT1beta}
\vspace{-0.3cm}
\end{figure}

To investigate the threshold behavior for lasers with $\beta \neq 0$, we examine the derivatives of $S(J)$~\cite{Barnes1976:IEEEJQE, Paoli1976:APL}.
The first derivative, $S'(J)$, grows gradually in the transition region, 
reflecting the increasing rate of the photon density. However, by obtaining the second derivative
\begin{widetext}
\begin{eqnarray}
S''(J)  = \frac{2g_0^2 n_T (1-\beta )\beta \Gamma^3 \tau_{ph}^3 \tau_r }{\left[4g_0 \beta \Gamma \tau_{ph} \tau_r J + \left( \beta- g_0 n_T \beta \Gamma \tau_{ph} +g_0  \Gamma \tau_{ph}(n_T- \tau_r J) \right)^2   \right]^{3/2}}, \label{Eq:dS0}
\end{eqnarray}
\end{widetext}
we can identify a characteristic single peak as a function of $J$. This is shown in the inset of Fig.~\ref{fig:JT1beta} for $\beta=0.01$. 
For a smaller $\beta$, the peak becomes sharper, and in the limit of $\beta\to0$, the peak becomes a Dirac $\delta$-function.
With $\beta \neq 0$, the lasing transition occurs at a range of injection centered at the peak. The peak position of $S''(J)$ represents the fastest change in the increasing rate $S'(J)$. Therefore, for $\beta\neq 0$, the injection corresponding to the peak of $S''(J)$ can serve 
as the representative value of the lasing threshold, which is given by 
\begin{eqnarray}
J_T(\beta) = n_T/\tau_r -\beta (2 n_T -n_{tran})/ \tau_r,
\label{Eq:JTbeta}
\end{eqnarray}
showing a linear decrease with $\beta$, as indicated by the vertical lines in Fig.~\ref{fig:1} for a few cases and specifically given by the dashed line in Fig.~\ref{fig:JT1beta}. We note that if 
\begin{equation}
\beta \geq 1/(2  -n_{tran}/n_T), 
\label{Eq:less}
\end{equation}
the peak feature of $S''(J)$ would be lost and therefore the threshold vanishes, i.e., the laser becomes thresholdless in this definition.  In Figs.~\ref{fig:1} and \ref{fig:JT1beta}, the injection is normalized by its threshold $J_T$ at $\beta=0$, and the photon densities are normalized by 
$S_{T}=S(2J_T)=\Gamma n_T \tau_{ph} /\tau_r$.  Unless otherwise specified, the
parameters in our calculations are: $\Gamma=0.02$, $n_{tran}=3.0\times 10^{12} $ cm$^{-2}$, $\tau_r=200$ ps, $\tau_{ph}=0.5$ ps~\cite{Westbergh2013:SPIE}, 
$\tau_s=200 $ ps, $g_0=60$ cm$^2$s$^{-1}$.

While our definition of a threshold relies on the experimentally measurable kink in the light-injection curves, 
other choices also exist in conventional lasers ($P=0$). For example, one used in microcavity lasers assumes that $S$ represents the density inside of the the gain region 
of a volume $V$ and that at threshold the photon number in the lasing mode is  $SV=1$~\cite{Yu:2003}.  Both definitions share similarities, $J_T(\beta \neq 0) < J_T(\beta= 0)$,
but in our case $J_T$ is linearly reduced with $\beta$ (Fig.~\ref{fig:JT1beta}).  

With various experiments on nano and microcavity spin lasers~\cite{Cheng2014:NN,Gerhardt:private}, in which $\beta$ exceeds previous values by several orders
of magnitude~\cite{Rudolph2005:APL,Holub2007:PRL}, there is an outstanding question about the influence of large spontaneous emission when $P \neq 0$. 
Due to the optical selection rules, light with different helicities $S^{\pm}$ exhibit separate thresholds $J_{T1, 2}$~\cite{Holub2007:PRL,Iba2011:APL} 
under spin injection, as shown in Fig.~\ref{fig:1}. 
Ignoring spontaneous emission ($\beta=0$), the first threshold of spin lasers can been obtained as~\cite{Gothgen2008:APL}
\begin{eqnarray}
J_{T1} (P) =  \frac{1}{1 +|P|/(4\tau_r/\tau_s + 2)} J_T,
\label{Eq:TR1}
\end{eqnarray}
where the minimum threshold $J_{T1}^{min} =  (4\tau_r + 2 \tau_s)/(4 \tau_r+3\tau_s) J_T$ is reached at $|P|=1$, which gives $J_{T1}^{min} /J_T=  2/3 $ in the limit of long spin-relaxation 
time ($\tau_s\gg \tau_r$)~\cite{Iba2011:APL,Iba2020:PSPIE}. For the
recombination linear in $n$,  threshold reduction is  up to $1/3$, smaller than the widely accepted theoretical limit 
$1/2$~\cite{Rudolph2003:APL, Rudolph2005:APL, Holub2007:PRL, Hovel2008:APL}. This is because the spin-dependent optical gain contains contribution from the spin-polarization of holes, the threshold reduction of $1/2$ is attained for their infinite spin-relaxation time. However, consistent with typical III-V semiconductors~\cite{Zutic2004:RMP}, hole spin-relaxation times are much shorter than for electrons, and therefore $p_+=p_-$ is assumed in Eqs.~(\ref{eq:ren}) and (\ref{eq:reS}), which diminishes the maximum threshold reduction. 

Analytical solutions of the steady-state rate equations with spin injection are generally unavailable. However, in the regime of a 
single-mode lasing $J_{T1}<J<J_{T2}$, the photon density of the nonlasing mode with spin-disfavored helicity is relatively small, e.g., $S^-\ll S^+$. Therefore, assuming $S^-\approx 0$ is accurate and simplifies the steady-state equations. Under this 
approximation, the analytical solutions for $S^+ (J) $ and $S''(J)$ can be obtained, which are qualitatively the same as for conventional lasers [see Eq.~(\ref{Eq:dS0})], only with more complicated expressions. 
Therefore, we propose that
the same definition of $J_{T1}$ as the peak position of $S''(J)$ developed in conventional lasers can be extended to spin lasers. Despite the  complexity of the full expression of $J_{T1}$, it is instructive to look at its simplification in the limit of long spin-relaxation times ($\tau_s \gg \tau_r$)
\begin{widetext}
\begin{eqnarray}
J_{T1}(P, \beta)=\frac{ 4+2|P|-(7 +5|P|)\beta +\Gamma g_0 \tau_{ph} n_{tran} (1-\beta) [4  +2|P| -(1-|P|) \beta]}{\Gamma g_0 \tau_{ph} \tau_r [2+ |P|-(1-|P|) \beta/2]^2 },
\label{eq:JT1beta}
\end{eqnarray}
\end{widetext}
which reduces to Eq.~(\ref{Eq:TR1}) for $\beta=0$ in the limit of $\tau_s \gg \tau_r$. 

To illustrate the effect of spontaneous emission and polarization of injection, we show in Fig.~\ref{fig:JT1beta} the dependence of $J_{T1}$ on $\beta$ for  $|P|=0$, $ 0.5$, and $0.9$. The result %
for the conventional laser ($|P|=0$) is based on Eq.~(\ref{Eq:JTbeta}). We see that for all cases, the lasing thresholds decrease approximately linearly with $\beta$, while $|P| \neq 0$ leads 
to a further threshold reduction. The decreasing rates for the spin-unpolarized and spin-polarized lasers are different. We note that the circular polarization of the emission would be reduced with increasing $\beta$ due to a larger contribution from spontaneous emission, which could be partially compensated by a longer spin-relaxation time. 

\begin{figure}[ht]
\centering
\includegraphics*[width=8.6cm]{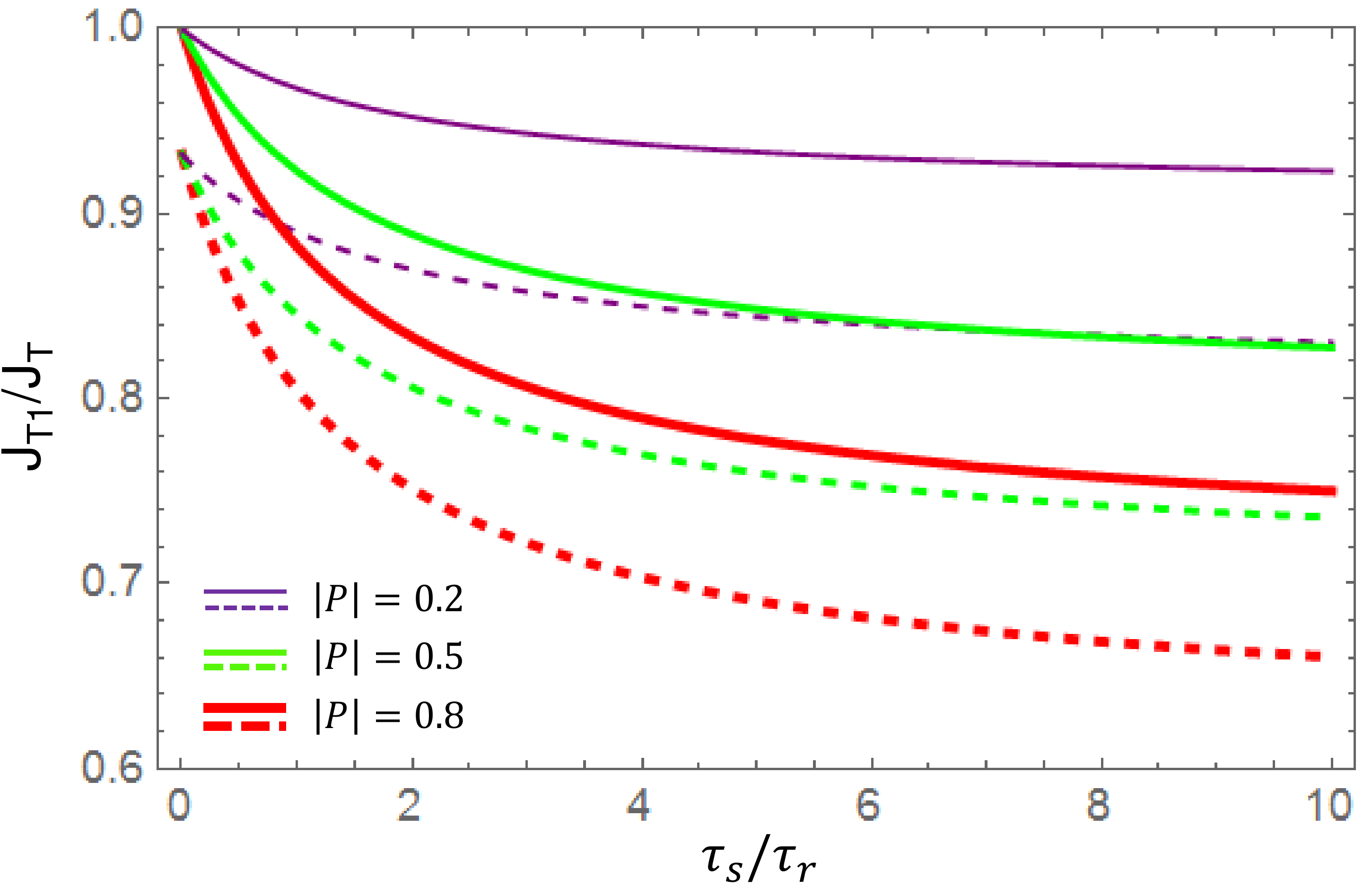}
\vspace{-0.5cm}
\caption{Influence of the normalized spin relaxation time $\tau_s/\tau_r$ on the normalized threshold $J_{T1}/J_T$ for $\beta=0$ (solid), $0.1$ (dashed) and $|P|=0.2, 0.5, 0.8$. }
\label{fig:tau_s}
\vspace{-0.5cm}
\end{figure}

While a moderate spin relaxation time $\tau_s=\tau_r = 0.2$ ns is used in Fig.~\ref{fig:JT1beta}, the threshold of spin lasers can be significantly influenced by $\tau_s$, as 
illustrated in Fig.~\ref{fig:tau_s}, where the corresponding dependence of the threshold $J_{T1}$ for $\beta=0$ and $0.1$ is shown. 
The curves show a similar trend that, with small $\tau_s$, $J_{T1}$ approaches the threshold value of  the corresponding 
conventional laser, while threshold reduction increases with $\tau_s$.  
This allows us to establish a steady-state picture and key trends of the threshold behavior of (spin) 
lasers, by combining our results for conventional and spin lasers with $\beta=0$ and $\beta \neq 0$. While the illustration here is given for $\beta$ up to $0.1$, the observed trends are also retained outside of this range. 

\begin{figure}[ht]
\centering
\includegraphics*[width=8.6cm]{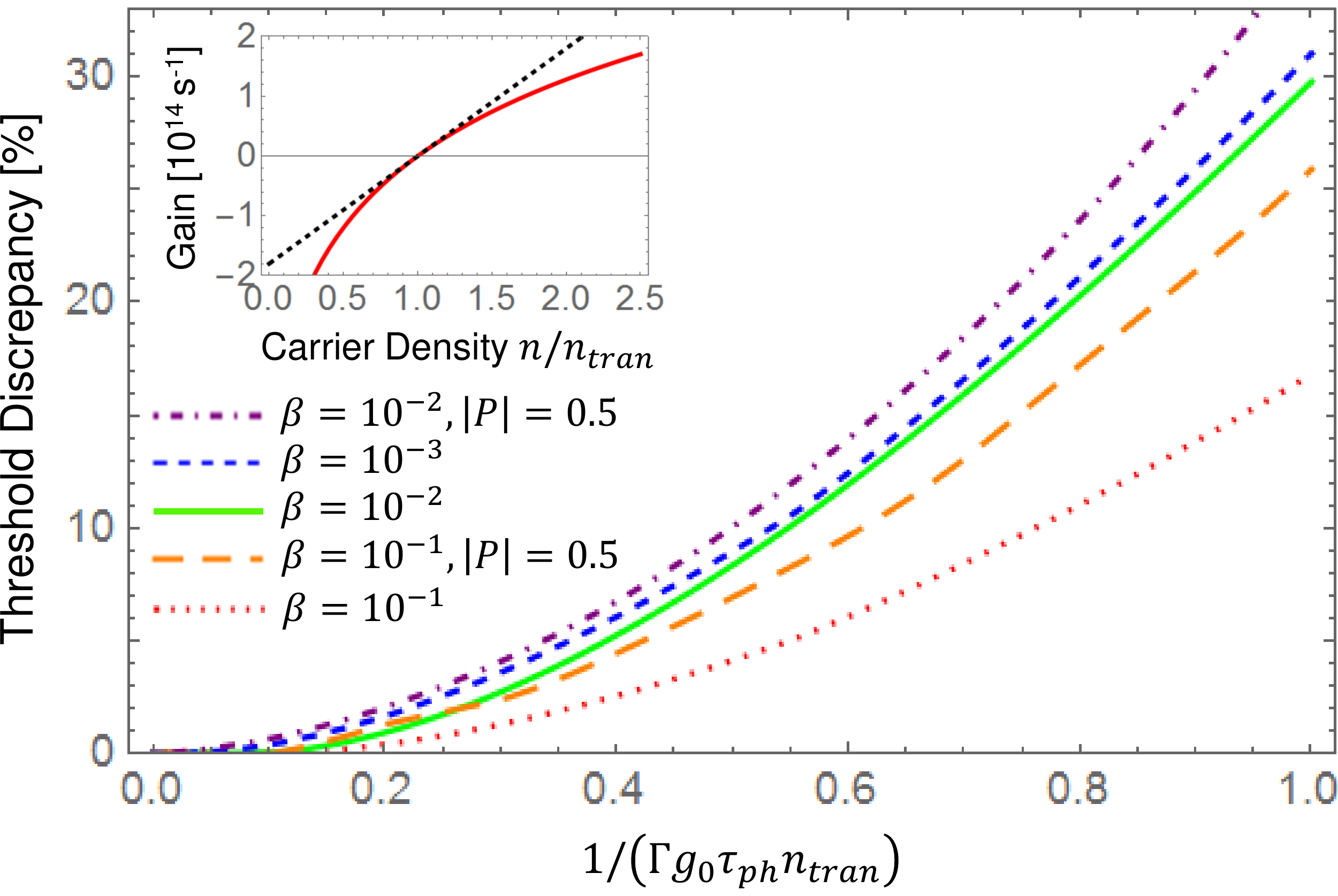}
\vspace{-0.5cm}
\caption{The relative discrepancies in thresholds between linear and logarithmic gain models as a function of $1/(\Gamma g_0 \tau_{ph} n_{tran})$ for $\beta=10^{-3}$, $10^{-2}$, $10^{-1}$ for conventional and spin lasers (with $|P|=0.5$). The inset shows a comparison of linear (dashed) and logarithmic gain curves (solid), with the carrier density normalized by $n_{tran}$. }  
\vspace{-0.3cm}
\label{fig:Log}
\end{figure}

So far, our rate equations have relied on a simple linear gain model, which is a good approximation in the vicinity of $n_{tran}$ and for bulk-like materials~\cite{Chuang:2009,Coldren:2012}.
However, the optical gain is generally nonlinear, especially for low-dimensional gain regions and carrier densities far from $n_{tran}$, where the accuracy of a linear gain model declines. In contrast, a logarithmic gain model can accurately describe the nonlinearity of realistic gain curves, with effects such as gain saturation 
accounted for~\cite{Chuang:2009,Coldren:2012,FariaJunior2017:PRB}. To describe the spin-dependent gain, we generalize a typical logarithmic gain 
model~\cite{Chuang:2009,Coldren:2012} to obtain
\begin{eqnarray}
g_{\pm}(n_{\pm},p_{\pm}) = g_0' n_{tran} \ln \frac{n_{\pm}+p_{\pm}+n_s}{n_{tran}+n_s},
\label{eq:LogGain}
\end{eqnarray}
where $n_s$ is a fitting parameter to be determined by the specific gain curve. For a better comparison, we make a connection between the gain models by matching the differential gain at $n_{tran}$, which leads to 
$g_0'=\left(1+ n_s/n_{tran}\right) g_0$. 
Here we set $n_s = 0.1 n_{tran}$ for the purpose of illustration. A comparison of linear and logarithmic gain curves is shown in the inset of Fig.~\ref{fig:Log}, where the two curves coincide at $n_{tran}$ and deviate  from each other away from $n_{tran}$. 

For $\beta=0$, the steady-state solutions with the logarithmic gain for  conventional lasers are
\begin{eqnarray}
n_T^{ln}&=&(n_{tran}+n_s) \exp [1/(\Gamma g_0' n_{tran} \tau_{ph})] - n_s   , \label{Sol:SteadyLog1} \\
S^{ln}&=& \Gamma \tau_{ph} (J-n_T^{ln}/\tau_r),  \label{sol:SteadyLog2}
\end{eqnarray}
which leads to the threshold injection $J_T^{ln}= n_T^{ln}/\tau_r$. Therefore, the threshold difference between logarithmic and linear gain is determined by the corresponding threshold carrier densities:
$J_T^{ln}-J_T^{lin}= (n_T^{ln}-n_T^{lin})/\tau_r$. 
An intuition can be gained by considering the regime $(n_T^{ln}-n_{tran})/n_{tran}\ll1$, i.e., $1/(\Gamma g_0' n_{tran} \tau_{ph})\ll1$, which allows an expansion of the exponential function in Eq.~(\ref{Sol:SteadyLog1}). The leading order approximation gives
$n_T^{ln}-n_T^{lin}  \approx  (n_{tran} + n_s)/[2(\Gamma g_0' n_{tran} \tau_{ph})^2]$, 
which suggests that in general $n_T^{ln}>n_T^{lin}$ and thus $J_T^{ln}>J_T^{lin}$,  with the difference largely dependent on the factor $1/(\Gamma g_0' n_{tran} \tau_{ph})$. This can be understood since the logarithmic gain includes the gain saturation, and thus it has a lower gain value at the same carrier density above $n_{tran}$ (see the inset of Fig.~\ref{fig:Log}). Therefore, a larger $n$  
and thus $J$ is needed to reach the lasing threshold. 

Unlike the case of the linear gain, an analytical steady-state solution of $S(J)$ with $\beta \neq 0$ is no longer available due to the complexity 
of the logarithmic function. However, our analysis for the case of $\beta=0$ that the threshold discrepancy between linear and logarithmic gain models depends on the threshold carrier density can be extended. For simplicity and connection with linear gain models, we consider the dependence of the discrepancy on  
$(n_T^{lin}-n_{tran})/n_{tran}=1/(\Gamma g_0\tau_{ph} n_{tran}) $ instead. 

We show in Fig.~\ref{fig:Log} the relative discrepancy of threshold between logarithmic and linear gain $(J_T^{ln}-J_T^{lin})/J_T^{lin}$ as a function of $1/(\Gamma g_0 \tau_{ph} n_{tran})$ for $\beta=10^{-3}, 10^{-2}$, and $10^{-1}$ for conventional and spin lasers (with polarization of injection $|P|=0.5$). For all cases, the discrepancy approaches to $0$ in the limit of small $1/(\Gamma g_0\tau_{ph} n_{tran})$. The relative discrepancy increases monotonically with $1/(\Gamma g_0\tau_{ph} n_{tran})$ and it is smaller with larger $\beta$. Moreover, the relative discrepancy in $J_{T1}$ of spin lasers is greater than that of the respective conventional lasers. Therefore, $1/(\Gamma g_0\tau_{ph} n_{tran})$ is a characteristic quantity 
that indicates the accuracy of the linear gain model in calculating the lasing threshold.  If $1/(\Gamma g_0\tau_{ph} n_{tran})  \ll 1$, linear gain models agrees well with logarithmic models. Otherwise, their discrepancy becomes significant. A simple linear gain model often underestimates the lasing threshold due to the neglect of the gain saturation, and its accuracy can be inferred from  the magnitude of the quantity $1/(\Gamma g_0\tau_{ph} n_{tran})$. 

While we have focused here on the steady-state response of spin lasers, our findings have broader implications and could guide the design of future scaled-down lasers,
where both large spontaneous emission and nonlinear gain are expected to play important roles. For example, in the limit $\beta=0$, it was found that  threshold reduction
also results in an enhanced modulation bandwidth~\cite{Lee2010:APL}, suggesting that with a large $\beta$ a further threshold reduction could be desirable. 
 
There is a growing interest to utilize optical anisotropy in spin lasers, such as the high birefringence~\cite{FariaJunior2015:PRB,Lindemann2016:APL,Yokota2018:APL,Pusch2017:APL,Pusch2019:EL,Pusch2015:EL,Lindemann2020:AIPA,Fordos2017:PRA,
Yokota2021:MM,Drong2021:PRA}, to reach ultrafast operation, faster than the best
conventional lasers~\cite{Lindemann2019:N}. Even though such efforts focus on quantum-well based lasers (where a nonlinear optical gain is expected) and scaling them 
down~\cite{Gerhardt:private},
their current description relies on a widely-used spin-flip model~\cite{SanMiguel1995:PRA,Al-Seyab2011:IEEEPJ,Adams2018:SST,Drong2020:JO}. 
While that approach includes optical anisotropies, the assumed linear gain and $\beta=0$ 
limits its predictive power to elucidate scaled-down spin lasers. Alternatively, optical anisotropy can be tailored in vertical external cavity surface emitting lasers, 
which are another promising platform to  implement spin lasers~\cite{Frougier2013:APL,Frougier2015:OE,Alouini2018:OE}, where it would also be important to study the role of a nonlinear gain and large $\beta$.

Beyond the spin lasers, the question of the threshold behavior continues to also be debated in conventional counterparts. 
A growing class of atomically-thin materials, such as transition metal dichalcogenides, were suggested as promising for spin lasers~\cite{Lee2014:APL} and later used
as the gain region in conventional lasers with a large $\beta$~\cite{Ye2015:NP,Wu2015:N}. However, within analyzing a linear optical gain model, these experiments were argued not
to have achieved a lasing threshold~\cite{Javerzac-Galy2018:NL}, prompting a need for both additional experiments as well as further theoretical analysis which may benefit from our findings. 

This work has been supported by the NSF ECCS-1810266 and 2130845. We thank N. C. Gerhardt for valuable discussions.

\section{Author Declarations}

\subsection{Conflict of interest}
The authors have no conflicts to disclose. 

\section{Data Availability}
The data that supports the findings of this study are available within the article.

\end{document}